\documentclass[conference]{IEEEtran}
\IEEEoverridecommandlockouts
\usepackage{cite}
\usepackage{amsmath,amssymb,amsfonts}
\usepackage{algorithmic}
\usepackage{graphicx}
\usepackage{textcomp}
\usepackage{xcolor}
\def\BibTeX{{\rm B\kern-.05em{\sc i\kern-.025em b}\kern-.08em
    T\kern-.1667em\lower.7ex\hbox{E}\kern-.125emX}}
\begin{document}

\title{Measuring a Low-Earth-Orbit Satellite Network}

\author{Jianping Pan, Jinwei Zhao and Lin Cai\\University of Victoria, BC, Canada}

\maketitle

\begin{abstract}
Starlink and alike have attracted a lot of attention recently, however, the inner working of these low-earth-orbit (LEO) satellite networks is still largely unknown. This paper presents an ongoing measurement campaign focusing on Starlink, including its satellite access networks, gateway and point-of-presence structures, and backbone and Internet connections, revealing insights applicable to other LEO satellite providers. It also highlights the challenges and research opportunities of the integrated space-air-ground-aqua network envisioned by 6G mobile communication systems, and calls for a concerted community effort from practical and experimentation aspects.
\end{abstract}


\section{Introduction}

The rapid growth of Starlink~\cite{starlink}, including over 4,000 launched low-earth-orbit (LEO) satellites and the associated ground infrastructure, as well as over 1.5 million subscribers around the world currently, has attracted a lot of attention from the industry and research community~\cite{leoconn}. Competing efforts such as OneWeb and Kuiper are also underway. They promise to have a global coverage and reach under-served population with broadband Internet. They further revolutionize the landscape of traditional satellite communication (SatCom) systems. However, the inner working of Starlink and alike is still largely unknown, and Starlink itself is also a moving target with continuous improvement and more satellites.

Compared to traditional SatCom systems, Starlink and alike employ many more LEO satellites much closer to ground users with significantly lower propagation delay, and offer much more capacity. On the other hand, both user terminal (UT) and ground station (GS) have to switch between LEO satellites constantly to maintain connectivity. In addition, Starlink will eventually use inter-satellite link (ISL) throughout its entire system to fully take the speed-of-light advantage in space over optical fibers. It provides a rare opportunity to communication and networking industry and research communities for the integrated space-air-ground-aqua (SAGA) network currently envisioned by 6G mobile communication systems.

This paper presents an ongoing measurement campaign focusing on Starlink, as the only operational LEO SatCom system with a massive user base now. Starting with a few UTs, we measured Starlink access networks, gateway and point-of-presence (PoP) structures, and backbone and Internet connections, leveraging the viewpoints and assistance of RIPE Atlas probes~\cite{atlas} and Reddit Starlink enthusiasts~\cite{reddit}, as well as Starlink's regulatory filings~\cite{nsf}. The revealed insights not only show the difference from traditional SatCom and conventional terrestrial Internet service providers (ISPs) but also apply to other LEO satellite providers. This can help the research community envision the 6G SAGA networks better.

More importantly, this paper highlights the challenges and research opportunities of SAGA networks and calls for a concerted community effort beyond RIPE Atlas, given the geographical coverage and diversity of Starlink and alike. With more community involvement from different locations around the world, the research community can have a better and more accurate understanding of Starlink and other LEO systems. Unlike the traditional SatCom systems with a few giant geosynchronous equatorial orbit (GEO) satellites and GSs and the conventional terrestrial ISPs mainly with regional coverage, Starlink and alike pose their unique challenges deserving global attention right now.

The rest of the paper is organized as follows. In Section~\ref{sec:background}, we briefly introduce Starlink including its user equipment and service offerings and related research efforts. After user-perceived performance metrics such as throughput, delay and loss with different UT and service prioritization, we present the Starlink access networks in Section~\ref{sec:access}, gateway and PoP structures in Section~\ref{sec:pop}, and backbone and Internet connections in Section~\ref{sec:backbone} with detailed topology diagrams first discovered by us, followed by research challenges, opportunities and community efforts in Section~\ref{sec:discussion}. Section~\ref{sec:conclusion} concludes the paper with further remarks.

\section{Starlink Background and Related Work}
\label{sec:background}

\subsection{Starlink in a Nutshell}
\label{sec:starlink}

Starlink~\cite{starlink} is a LEO satellite network for broadband Internet access and backbone, put together by SpaceX with its revolutionary reusable rocket launch technology and commodity satellite manufacturing process, which greatly reduced the cost, complexity and turnaround time of building and maintaining an operational SatCom system. Currently Starlink has more than 4,000 LEO satellites in different generations launched into orbit in different shells (inclination and altitude combinations), with several thousands more approved and to be launched, but the regulatory filings are still changing and adapted to the need of SpaceX. Most current satellites are in 53$^\circ$ inclination at 550km above the Earth.

In addition to satellite telemetry, tracking and control (TTC) channels, each satellite has a number of UT and GS-facing phased-array antennas in Ku and Ka bands, respectively, with E bands added to new generations and cellular bands to future ones. The specific frequency and transmission power are regulated by different countries at different locations. Similar to traditional SatCom systems, these frequencies are susceptible to atmospheric impairments and obstacles including heavy rain, snow and trees. Both UT and GS antennas (commonly known as ``flat'' and ``dome'' dish, respectively) need a clear view of the sky with 25$^\circ$ minimal elevation above the horizon.

Starlink subscribers receive a self-installation kit with different mounting options. The first-generation ``round'' dish has similar performance as the second-generation rectangular dish (``v3'' currently), albeit a larger surface and more antenna elements as the latter has a higher efficiency. Round and v3 dishes are widely used for standard/residential services. The high-performance ``HP'' square dish has more power and higher efficiency, and is used for priority/business, maritime and in-motion mobility/roaming services, with more portable dishes upcoming. Starlink subscribers pay for their dish upfront, and their monthly fee determined by the service subscription, without long-term contracts or hard data caps.

Once Starlink subscribers installed the dish, they can use Starlink-provided router with WiFi, or use their own router through the Ethernet port or adapter, to connect their devices to the Internet, just like other forms of Internet access, where the router provides the network address translation (NAT) functionality to allow multiple user devices connected to the Internet with a public IP address. For Starlink business users, the public IP address is static and accessible from the Internet, while other types of Starlink users share public IP addresses through Starlink's carrier-grade NAT (CGNAT) at the GS, where the public IP address reflects the associated PoP.

\subsection{Related Work}
\label{sec:related}

Research on SatCom has a long history, but mostly focused on GEO satellites. Even before Starlink was launched, the new approach to LEO SatCom has attracted a lot of attention from the research community. With regulatory filings, various geometry-based simulation appeared with a back-of-the-envelop calculation of Starlink system capacity and user performance, some assuming ISL and ground or vessel relay, etc~\cite{space-routing,ground-relay,space-race,topology-design}. Simulation-based network performance studies also emerged, as well as network emulation and testbed construction~\cite{hypatia,testbed,with-quic}. On the other hand, many enthusiastic Starlink users, some from the initial beta test program, reported their experience on Reddit, despite some inconsistency~\cite{reddit}.

One notable network measurement effort is enabled by RIPE Atlas~\cite{atlas}, with more than 10 thousand active probes and anchors deployed by Internet users and ISPs around the world, actively measuring network liveliness using tools such as {\tt ping}, {\tt traceroute} and {\tt nslookup}. Currently, there are about 65 active probes in about 17 Starlink access networks, located in the USA, Canada, Australia, New Zealand, and some European, Asian and South American countries. However, compared to the Starlink global coverage so far in more than 50 countries, there are still considerable regions without any active Atlas probe. One purpose of this paper is to promote more community members to host Atlas probes.

The most related work is a measurement study to a few Starlink dishes from cloud data centers around the world, and the results are thus also dominated by the global Internet~\cite{from-cloud}. In this work, we have access to a few dishes while also leveraging Atlas probes and Reddit assistance. In particular, we focus on the Starlink access networks with different dishes and subscription tiers, the gateway and PoP structures at different locations, and the backbone and Internet connections with other ISPs. Through network measurement for almost a year, we present the first detailed topology diagrams of Starlink access, PoP and backbone networks after crosschecked with Atlas and Reddit communities with high confidence.

\section{Starlink Network Measurement}

We acquired several Starlink dishes and gained access to a dozen more through the assistance of Starlink users, initially associated with the Seattle PoP and now expanding to other PoPs, complemented by the ability of RIPE Atlas probes and the data provided by Reddit users around the world, so we can measure the Starlink system from multiple vantage points with different dishes and service tiers. The results are then validated by more Reddit Starlink users, representing the state-of-the-art understanding of Starlink other than by SpaceX itself.

\subsection{Starlink Access Networks}
\label{sec:access}

Starlink access networks include the user network facilitated by the Starlink-provided or user-provisioned router, connecting multiple user devices to the Internet through a NAT on the router. The router is connected to the Starlink dish, collectively known as Starlink UT, through a power-over-Ethernet (PoE) cable. The user dish communicates with ground dish, through one or possibly multiple satellites. Users are grouped into service cells, roughly 24km in diameter, and a rule-of-thumb capacity limit now is ``100 users in every 300 km$^2$.''

\subsubsection{Access Topology and Structure}

\begin{figure}
    \centering
    \includegraphics[width=\columnwidth]{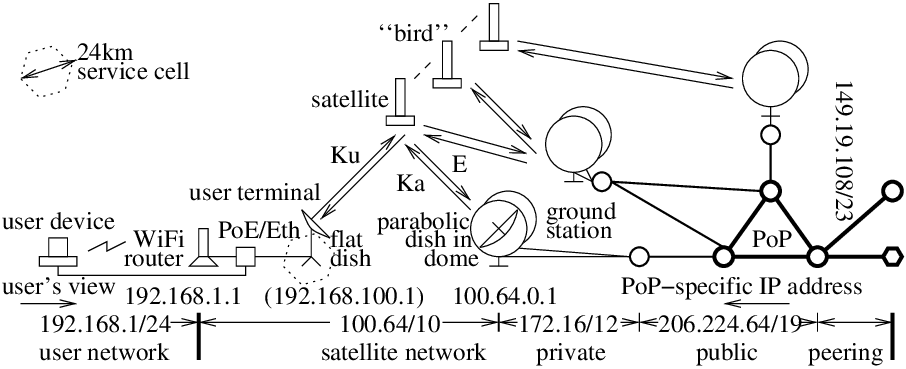}
    \caption{Measured Starlink access networks near Seattle (similarly elsewhere).}
    \label{fig:access}
\end{figure}

The access network topology and structure is shown in Fig.~\ref{fig:access}. Starlink-provided router is fixed at 192.168.1.1 (user-provisioned router can use other private address), and the Starlink dish is always at 192.168.100.1, from the user's viewpoint, so users may need to add a static route to reach the dish if using their own router. From Starlink side, user router will have a CGNAT address in 100.64/10 and Starlink can remotely access Starlink-provided routers with consent. Starlink dish shall have a 100.64/10 address too, although it has not been independently verified as the communication between the dish and its gateway is inaccessible now and claimed to be encrypted. 

Starlink satellites, also known as ``birds'', have no user visibility at IP layer. Not all satellites are now ISL capable or have ISL enabled, which can be observed through the delay between user and ground dishes. Each satellite has multiple antennas subdivided into beams. According to a patent granted to SpaceX~\cite{mac}, Starlink uses a very simple grouping-based media access control (MAC) scheme at the satellite for UT and GS links. Normally four users share a communication channel, and the satellite polls user dishes periodically to grant access for the uplink. With larger propagation delay due to the distance and considerable MAC delay due to the polling, the minimum Starlink access round-trip time (RTT) is about 20ms.

The ground dish is always at 100.64.0.1 from user's viewpoint, which is also the CGNAT, i.e., translating user's 100.64/10 address to a public IPv4 address. For business users, the public address is statically bounded at their router and also reachable from the Internet, and for other users, it is shared and temporary, and user router cannot be reached directly unless proper NAT traversal is done at CGNAT, e.g., through {\tt tailscale}. With IPv6, Starlink users are better reached but not all networks are IPv6 operational now, so we only focus on IPv4 in this paper. The public address reflects the location of users and their subscription, specific to the associated PoP, for geo-location and third-party service provisioning.

\subsubsection{Access Measurement and Performance}

\begin{figure}
    \centering
    \begin{tabular}{cc}
       \includegraphics[width=0.47\columnwidth]{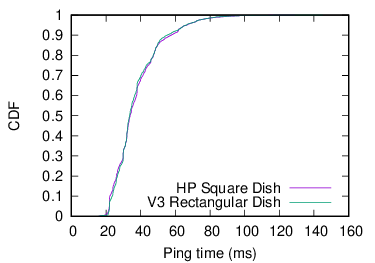} & \includegraphics[width=0.47\columnwidth]{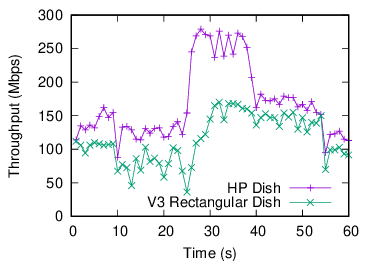}\\
       \includegraphics[width=0.47\columnwidth]{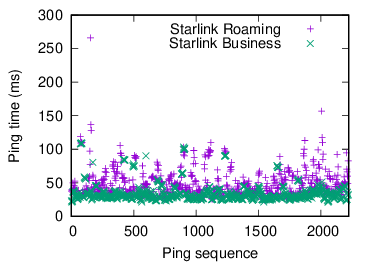} & \includegraphics[width=0.47\columnwidth]{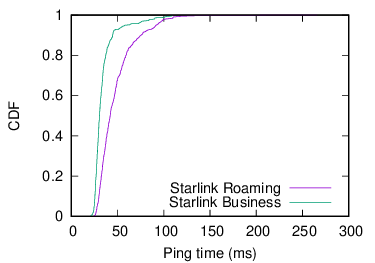}
    \end{tabular}
    \caption{Measured Starlink access performance near Seattle (c/o Nathan).}
    \label{fig:access-performance}
\end{figure}

Figure~\ref{fig:access-performance} shows the latency and throughput performance comparison of Starlink dishes and service tiers. HP and v3 dishes have similar latency performance to 100.64.0.1 for residential users, but HP dish can have a higher throughput to an {\tt iPerf3} server peered at the associated PoP for both downlink and uplink (not shown, 14.3 vs 6.94Mbps) due to more antenna elements, higher transmission power and better capability to track satellites. On the other hand, business users, which also use HP dishes, have considerably lower latency than roaming users, who similar to best-effort and portability users have the lowest priority when compared with residential, business and maritime users.

\begin{figure}
    \centering
    \begin{tabular}{cc}
       \includegraphics[width=0.47\columnwidth]{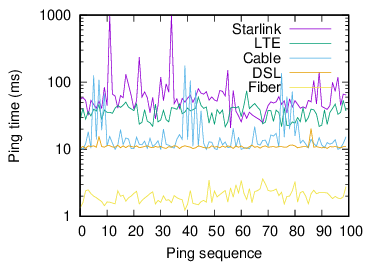} & \includegraphics[width=0.47\columnwidth]{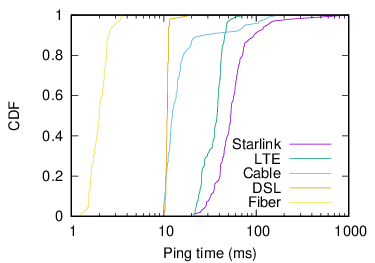}\\
       \includegraphics[width=0.47\columnwidth]{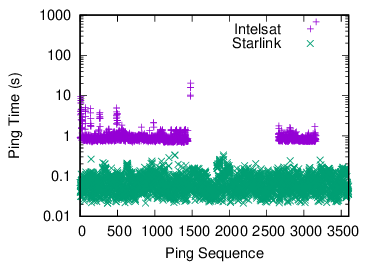} & \includegraphics[width=0.47\columnwidth]{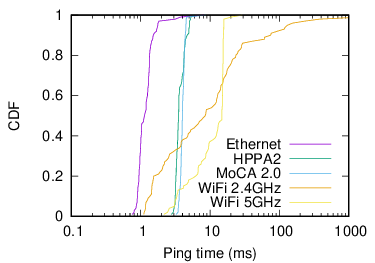}
    \end{tabular}
    \caption{Starlink access performance compared with other technologies.}
    \label{fig:access-comparison}
\end{figure}

The ping time sequence shows that the RTT to the gateway is highly fluctuating, reflecting the nature of satellite handover and competing users sharing the same beam. Comparing with other Internet access and home WiFi technologies as shown in Fig.~\ref{fig:access-comparison}, Starlink access RTT is indeed higher than fiber optics, digital subscriber line (DSL) and cable modem (although the cable uplink can be quite bursty due to its shared neighborhood), but comparable to long-term evolution (LTE) cellular and significantly better than traditional GEO SatCom such as Intelsat. When ISL is involved, the access RTT shows considerable stage effect, possibly going through a different number of satellites and different gateways to the same PoP. Starlink currently has no roaming support at the PoP level.

Please note that Starlink access is highly influenced by weather and traffic conditions. Users can use Starlink mobile app (or a Web browser pointing to 192.168.100.1) to know the outage events and statistics of their Starlink dish: obstructed (locally at the user dish), no signal received (between the user dish and satellite) and network issue (between the satellite and ground station), as well as ping time to their associated PoP (not the gateway as we measured) and observed uplink and downlink throughput (not capacity). Using gRPC tools, Starlink users can automate the process with many user-contributed monitoring and dashboard tools, including regular speedtest to popular sites, available on GitHub~\cite{grpc}.

\subsection{Starlink Gateway and PoP Networks}
\label{sec:pop}

Starlink gateways, loosely referring to the ground station dishes, antennas and beams, are connected to a regional or country-wide PoP. For some regions, Starlink users can be associated with neighbor PoPs, as reflected by their entries in the Starlink geo-location database~\cite{geoip} and reverse domain name system (DNS) lookup, for redundancy and load-balancing purposes. Here we use the Seattle PoP as an example.

\subsubsection{PoP Topology and Structure}

\begin{figure}
    \centering
    \includegraphics[width=\columnwidth]{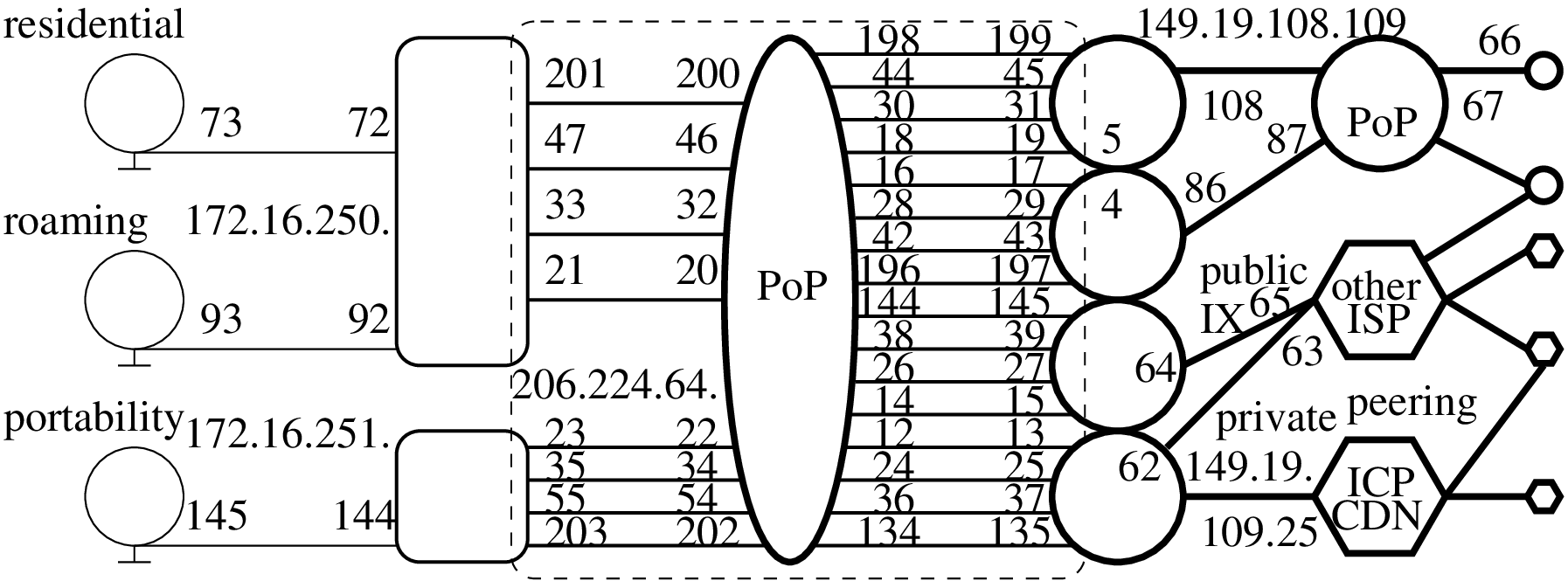}
    \caption{Measured Starlink gateway and PoP structures in Seattle (c/o Robert).}
    \label{fig:pop}
\end{figure}

Each gateway is identified by an address in 172.16/12 network, which is only unique in the same PoP, and different PoPs can reuse the same 172.16/12 address, similarly as any other private IP address space. Under Seattle PoP, we have observed gateways in 172.16.250/24, 172.16.251/24 and 172.16.252/24, where odd ending digits associated with the CGNAT and even ones associated with the  PoP. A Starlink user can reach all gateway identifiers within the same 172.16.x/24 (except those ending with 9) by ping, indicating they are associated with the same ground station, possibly different dishes or beams. However, the user cannot ping those in 172.16.y/24 in other ground stations.

The traceroute from Starlink users to 172.16.y/24 outside their own gateway station shows the packet reaches the PoP identified by 206.224.64/19 for large PoPs (e.g., Seattle) or 149.19.108/23 for smaller PoPs (e.g., Denver), indicating some hierarchical structures among PoPs. Note that Starlink does not limit traceroute to 172.16/12 properly within its PoP and may route toward the public Internet, incurring network unreachable messages returned by some routers. Starlink has been alerted on this private route leaking issue but no fixes have been observed yet. Traceroute to the public IP address of Starlink users depends on whether it traverses the CGNAT.

Starlink PoP structure is quite similar to other terrestrial and cellular ISPs, with leased fiber connections to ground stations usually located in rural areas or on hill or building tops for a better sky view. Each ground station has multiple (often nine) parabolic dishes in protective domes. For Seattle PoP, it covers ground stations even in Alaska, and for Los Angeles PoP, it covers Hawaii, indicating possible long-haul fiber connections between gateways and its associated PoP. This can cause unnecessarily long delay for local traffic within a PoP, as observed by many Starlink users and reported on Reddit, so Starlink can make further improvement.

\subsubsection{PoP Measurement and Results}

Figure~\ref{fig:pop} shows a snapshot of the Seattle PoP with three dishes and service tiers: residential with round dish, and roaming and portability with rectangular dish. Their public IP addresses change very often, reflecting the competition among users associated with the same PoP, but still indicate the service tier, e.g., residential and roaming near Vancouver, and portability from Seattle, all physically in Victoria. However, their gateway addresses are relatively stable but do change from time to time. At this snapshot, both residential and roaming are associate with the same ground station, and portability with another. Similar behaviors are observed in other PoPs in the US and elsewhere.

In large PoPs such as Seattle, there are two levels of interconnection facilitated by the 206.224.64/24 network. Different ground stations are interconnected by the PoP, which also peers with other PoPs, ISPs and content distribution networks (CDNs). Interconnection is organized in pods, and there are multiple parallel links between pods, which are revealed in Fig.~\ref{fig:pop}. TCP and ICMP messages are hashed to a particular link, while UDP packets can traverse parallel links, resulting in packet reordering and affecting some UDP-based applications. However, this is a common practice among ISP and Internet exchange provider (IXP), and is not unique to Starlink. 

Through our measurement, we observed different Starlink users can be assigned to the same gateway (172.16.x.z) but with different public IP addresses, or even the same public IP address for some time. One of them is a dish we have access and another is a different Atlas probe, so we can conclude that Starlink CGNAT can assign the same public IP address to different users at the same time, complicating user tracking. Starlink users reported on Reddit that they were served with Digital Millennium Copyright Act (DMCA) notices without involving in such activities, indicating Starlink mixed up user tracking due to excessive reuse of IP addresses, while keeping some addresses idle to switch GS between PoPs.

\subsection{Starlink Backbone and Internet Exchange Networks}
\label{sec:backbone}

Starlink runs its own backbone with global reach and interconnect with other ISPs around the world. Most access network providers do not have global coverage, while Starlink operates both its access and backbone networks, which is quite unique. Starlink's autonomous system (AS) number is 14593 and advertises many separate IP address blocks through BGP, which increases its routing complexity in today's world.

\subsubsection{Backbone Topology and Structure}

\begin{figure*}
    \centering
    \includegraphics[width=0.95\textwidth]{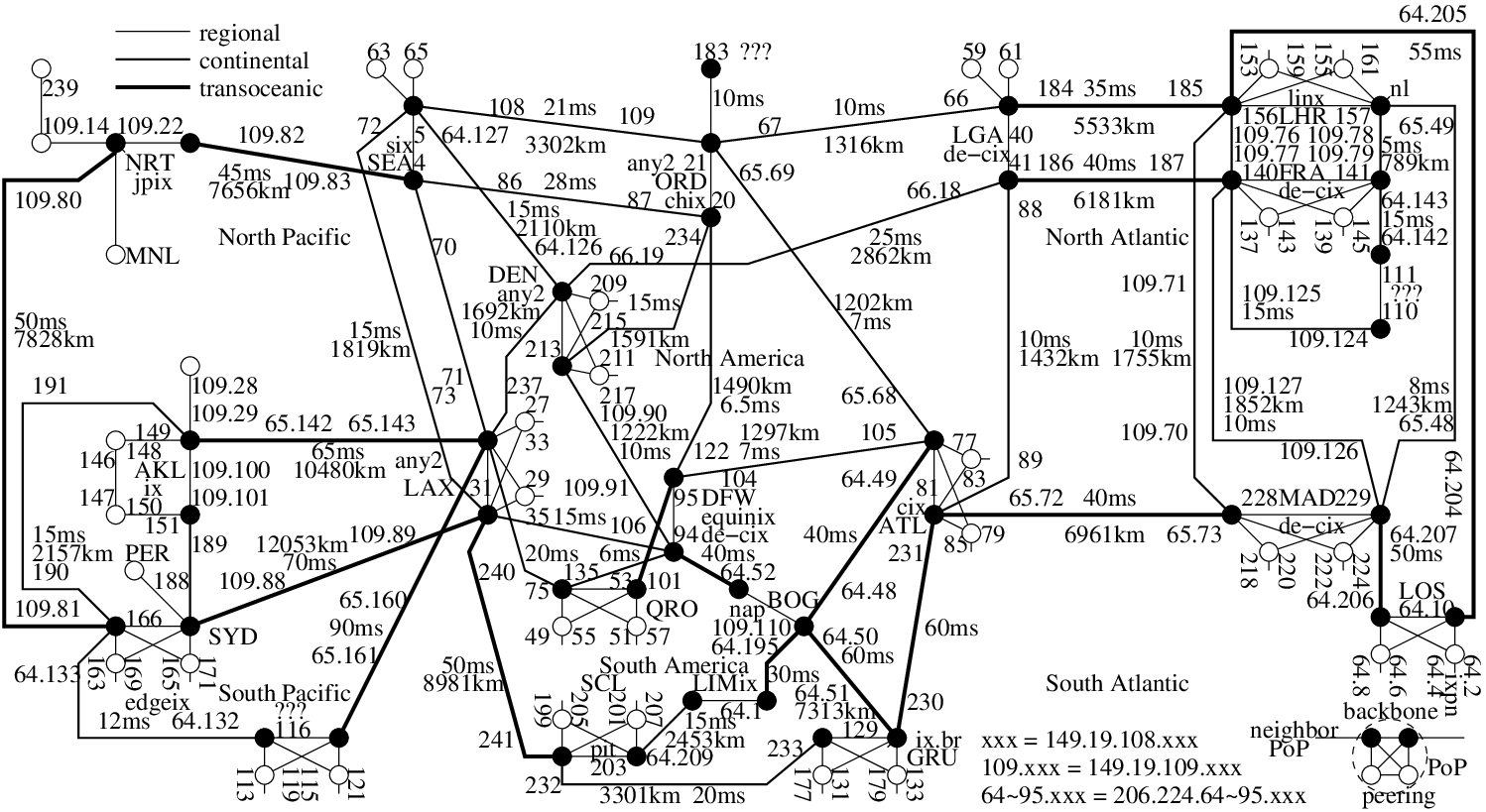}
    \caption{Measured Starlink backbone network around the world with Perth (PER) and Manila (MNL) PoP recently active in naming, addressing and routing.}
    \label{fig:backbone}
\end{figure*}

Most Starlink PoP has a very regular topology, with two interconnected backbone routers connected with at least two neighbor PoPs, and two cross-connected peering routers with other ISP and CDN at public or private IXP locations. Backbone addresses are in 149.19.108/24 initially, with 149.19.109/24 for peering. With the rapid growth of Starlink, 206.224.64/24, 206.224.65/24 and 206.224.66/24 are also used nowadays. In the USA, currently there are seven PoPs, namely Seattle (SEA), Los Angeles (LAX), Denver (DEN), Dallas (DFW), Chicago (ORD), Atlanta (ATL) and New York City (LGA), serving both American and Canadian customers, although there are a few GSs in Canada, mostly located on east coast (e.g., Newfoundland).

Worldwide, Starlink currently has three PoPs in Europe, i.e., London (LHR), Frankfurt (FRA) and Madrid (MAD), three in Oceania, i.e., Sydney (SYD), Auckland (AKL) and Perth (PER), two in Asia, i.e., Tokyo (NRT) and Manila (MNL), and one in Lagos (LOS), Africa. In addition, Starlink has one PoP in Queretaro (QRO), Mexico, and four more in South America, i.e., Bogota (BOG), Lima (LIM), Santiago de Chile (SCL) and Sao Paulo (GRU). The setup of PoPs certainly follows the local regulatory requirement, and different PoPs have very different range of ground station coverage. Starlink initially leveraged Google cloud platform (GCP) and later evolved to have its own backbone and direct peering agreements with other ISPs/IXPs.

\subsubsection{Backbone Measurement and Results}

To discover Starlink terrestrial backbone, we started with Starlink's entries in PeeringDB~\cite{peeringdb} and Starlink published GeoIP database~\cite{geoip}, where we can identify the public IP address blocks advertised by Starlink with their reverse DNS name lookup. Please note that Starlink's GeoIP database is not always updated and synchronized, as we found addresses advertised to be associated with Seattle PoP actually route toward Frankfurt, reflecting the fact that Starlink may move address blocks to serve the growing subscription in certain regions during certain events. Thus, we trace from known landmarks to determine the location of Starlink backbone addresses as shown in Fig.~\ref{fig:backbone}.

For example, the Seattle PoP has two interconnected backbone routers (149.19.108.4 and 5) connected to LAX, Denver and Chicago PoPs, and peers at {\tt six}, Seattle Internet Exchange. Using {\tt mtr} for a long time, we can determine the best (or minimum) RTT difference between neighbor hops, and treat it as the propagation delay since the transmission delay of ground links for ping messages is negligible. The half of the RTT difference, treated as one-way delay (OWD), is crosschecked with the travel distance between PoPs, as most fiber conduits are along highway or railway, which shows high correlation as highlighted in the figure in ms and km. This gives us the confidence that the discovered topology is close to reality, even not disclosed by Starlink yet.

Once we identified PoPs and backbone router addresses, we run systematic traceroute and when possible mtr to determine backbone links, where we also leveraged Atlas probes at different locations and received assistance from Starlink users who reported their service on Reddit. Starlink's backbone link numbering is quite regular. For example, 20--21 for Chicago PoP, 30--31 for LAX PoP, 40-41 for LGA PoP, 80--81 for Atlanta PoP, etc, although such regularity decays when Starlink expanded worldwide. For example, 160--161 is associated with London PoP and 162--163 with Sydney. Despite our best effort, so far we still cannot exactly locate clusters such as 116--117, 183 and 110--111 in 149.19.108/24 block.

We also used public looking glass, route and traceroute servers external to Starlink to verify the Starlink backbone, PoP, gateway and even access networks that we measured and discovered. So far, we have observed many BGP routing deficiencies. For example, traffic toward Starlink users does not always enter the nearest Starlink PoP. Although this is a common problem among other ISPs, the impact is more severe for satellite ISPs and their users, as they already suffer higher delay than those on terrestrial networks. On the other hand, Starlink has its own unique challenge as it is both a global backbone and access network provider. This requires more attention from the research community.

\section{Research Challenges and Opportunities}
\label{sec:discussion}

Although our measurement has revealed Starlink access, gateway, PoP and backbone networks in great details and also outlined some problems in the current Starlink network arrangement and provisioning, we found more challenges and opportunities deserving a community effort.

\subsection{Challenges of Integrated SAGA Networks}

Networking research has mainly focused on the Earth surface, although there are research efforts for the Moon and the interplanetary scenarios, but not at a consumer level. With the integrated SAGA networks envisioned by 6G, commercially viable SatCom becomes an immediate need. The reusable rocket launch technology enabled it, but the integration of networking protocols with TCP/IP is still lacking.

Specifically, most TCP/IP protocols have been designed and then optimized for terrestrial networks. Although there is link-layer mobility supported in mobile cellular  networks and WiFi, it is far behind what LEO satellites demand, where not only end users but also the network itself is mobile, although quite predictable given the physics. Currently, Starlink does not reveal its satellite relay scheme. Given the space and spectrum resource limitation, as well as geographical and even political constraints, LEO SatCom systems may have to inter-operate at certain levels around the world. This is a challenge that has not been tackled in ordinary cellular or WiFi systems.

Moreover, integrated SAGA also enables multiple paths, including those traversing satellites, between end hosts. Currently most Internet routing is single, shortest path routing by some metrics or policies, while the rich link and network diversity is ignored~\cite{percolation}. For SAGA networks, multi-path is not only an option to increase resiliency and load balancing but also a necessity for mission-critical tasks in deep space and distressed scenarios, as evidenced by the fact that more cellular phones adopted direct satellite SOS capabilities recently. Not only the shortest path but also maximal network flow becomes a need, especially in a distributed manner, given there are some recent breakthroughs in max-flow problems.

Last but not least, the dominant transport-layer protocol, TCP, still assumes packet loss due to network congestion and backs off in different ways, where in SAGA networks, packet loss and delay variation can come from different sources. Traditional GEO SatCom has engineered many TCP performance-enhancing proxies (PEPs) at the user terminal and ground station to improve TCP performance. Whether such PEPs are still beneficial in the LEO scenario needs to be evaluated again. On the other hand, new transport-layer protocol, such as QUIC, is emerging, with built-in connection migration capability, and its multi-path features are currently under active discussion at IETF for standardization.

\subsection{Opportunities through Starlink-like Systems}

Therefore, being the first LEO satellite network with global coverage and considerable user base, Starlink and its competitors in the near future are good testbed candidates for the research community. Recall that the current Internet was a testbed initially known as ARPANET. Starlink is not a perfect system, and it is also a moving target with more satellites added, services introduced and policies changed, on a weekly or monthly basis. However, it is a good opportunity for the research community to understand the SAGA network envisioned for 6G, just as the UCLA network measurement center (NMC) interacted with BBN who manufactured the IMP deployed on  ARPANET, which evolved into the Internet.

One rare opportunity is inter-satellite communication and networking. So far we have not had a dynamic network with very fast but regular topology changes such as LEO satellite networks, where the shortest-path and maximum-flow routing algorithms and protocols will have an ultimate test. TCP/IP was designed for a network with arbitrary topology thus unnecessary overhead when the network topology shows regularity either statically (e.g., in data centers) or dynamically like LEO satellites. It is the time to rethink TCP/IP.

Another opportunity is integrated sensing, communication and computing. With more powerful LEO satellites, not only satellite-to-ground and inter-satellite communications become commonplace, so do Earth-facing imaging and star-facing observation with less light and atmosphere pollution. However, not all sensed data can or need to be transferred to ground for processing, so on-board computing becomes a need. Amazon's Kuiper attempts to do so, leveraging its ground-based cloud-edge computing. Integrated SAGA opens a new dimension.

\subsection{An Appeal for a Global SAGA Testbed}

More importantly, we appeal to the research community, especially those who can acquire or access Starlink dishes, especially in remote locations, to host RIPE Atlas probes to help the research and also user community to better understand Starlink and similar systems. For research purposes, we also appeal to federate these dish-connected computers for remote access, just as what Planetlab (and later GENI) did at the start of this century for distributed systems and networking research. Experimentation is an important approach to initiate and validate our research and keep it practical and relevant.

\section{Conclusions}
\label{sec:conclusion}

In this paper, we presented an ongoing measurement campaign on Starlink, the first large-scale, operational, LEO satellite network, on its access, gateway, PoP and backbone networks with detailed network topology diagrams and some performance results. More importantly, we discussed the challenges and opportunities brought in by Starlink and alike, and appealed to the research community to create and participate in a global observatory testbed for a SAGA network as envisioned by 6G communication systems.


\begin{thebibliography}{00}
\bibitem{starlink} Wikipedia, Starlink, https://en.wikipedia.org/wiki/Starlink, 2023.

\bibitem{leoconn} LEOCONN, https://leoconn.github.io/, 2021 and 2022.

\bibitem{atlas} RIPE Atlas, https://atlas.ripe.net/, 2023.

\bibitem{reddit} Reddit, Starlink, https://www.reddit.com/r/Starlink/, 2023.

\bibitem{nsf} NSF, Starlink, https://forum.nasaspaceflight.com/index.php?topic=48981.0

\bibitem{space-routing} M.~Handley, ``Delay is not an option: Low latency routing in space,'' ACM HotNets, 2018.

\bibitem{ground-relay} M.~Handley, ``Using ground relays for low-latency wide-area routing in megaconstellations,'' ACM HotNets, 2019.

\bibitem{space-race} D.~Bhattacherjee, W.~Aqeel, I.~Bozkurt, A.~Aguirre, et al., ``Gearing up for the 21st century space race,'' ACM HotNets, 2018.

\bibitem{topology-design} D.~Bhattacherjee, and A.~Singla, ``Network topology design at 27,000 km/hour,'' ACM CoNEXT. 2019.

\bibitem{hypatia} S.~Kassing, D.~Bhattacherjee, A.~Aguas, J.~Saethre, and A.~Singla, ``Exploring the Internet from space with Hypatia,'' ACM IMC, 2020.

\bibitem{testbed} M.~Kassem, A.~Raman, D.~Perino, and N.~Sastry, ``A browser-side view of starlink connectivity,'' ACM IMC, 2022.

\bibitem{with-quic} F.~Michel, M.~Trevisan, D.~Giordano, and O.~Bonaventure, ``A first look at Starlink performance,'' ACM IMC, 2022.

\bibitem{from-cloud} S.~Ma, et al., ``Network characteristics of LEO satellite constellations: A Starlink-based measurement from end users,'' IEEE INFOCOM, 2023.

\bibitem{mac} J.~Iyer, K.~Mahammad, et al., ``System and method of providing a medium access control scheduler,'' US Patent 11,540,301, 2021.

\bibitem{grpc} Starlink gRPC Tools, https://github.com/sparky8512/starlink-grpc-tools

\bibitem{geoip} Starlink Self-Published IP Geo-location Feed, http://geoip.starlinkisp.net

\bibitem{peeringdb} PeeringDB, https://www.peeringdb.com/asn/14593, 2023.

\bibitem{percolation} J.~Hu, et al., ``Directed percolation routing for ultra-reliable and low-latency services in low earth orbit satellite networks,'' IEEE VTC-W'20.
\end{thebibliography}
\end{document}